\documentclass[prl,aps,twocolumn]{revtex4}
\usepackage{epsfig}
\usepackage{amsmath}
\begin{document}
\title{Viewing the Proton Through ``Color"-Filters}
\author{Xiangdong Ji}
\email{xji@physics.umd.edu}
\affiliation{Department of Physics,
University of Maryland,
College Park, Maryland 20742 }
\date{\today}
\begin{abstract}

While the form factors and parton distributions provide separately the
shape of the proton in coordinate and momentum spaces,
a more powerful imaging of the proton structure can be obtained through phase-space
distributions. Here we introduce the Wigner-type quark and gluon
distributions which depict a full-3D proton at every
fixed light-cone momentum, like what is seen through momentum(``color")-filters.
After appropriate phase-space reductions, the Wigner distributions are
related to the generalized parton distributions (GPD's) and
transverse-momentum dependent parton distributions, which are measurable
in high-energy experiments. The new interpretation of GPD's provides
a classical way to visualize the orbital motion of the quarks,
which is known to be the key to the spin and magnetic moment of the proton.

\end{abstract}
\maketitle

What is the shape of the proton?
The textbook answer is that it is round \cite{textbook}. Indeed, the proton is
a spin-1/2 particle, and the strong interactions are invariant
under time-reversal symmetry. Hence there cannot be any dipole, quadrupole or
higher-multiple deformations, and the quark and gluon density
distributions in space have to be spherically symmetric. It is well
known that the Fourier transformation of the proton's electric
form factor yields the {\it spherical} charge distribution when
the relativistic recoil effects are neglected \cite{textbook}.

The classical answer is correct, but might not be complete. The quarks
and gluons are not static inside the proton, they are moving relativistically
to produce the magnetic moment and part of the angular momentum
(spin) of the proton. The orbital motion has an axial, rather than spherical,
symmetry when the proton spin is polarized along a fixed direction \cite{sachs}.
To be sure,
the Feynman distributions do provide a momentum space
picture of partons, but there is no spatial density information involved---
they are just {\it correlation} functions in the position space. To gain
a more complete understanding of the proton's structure, it is useful to
know the quarks' and gluons' joint-position-and-momentum (or phase-space)
distributions. Indeed, in classical physics, a
state of a particle is specified by knowing both its position
$x$ and momentum $p$. In a gas of classical identical particles,
the single-particle properties are described by a phase-space
distribution $f(\vec{r},\vec{p})$ representing the density of particles
at phase-space positions $(\vec{r},\vec{p})$. Time-dependence
of the distribution is governed by the Boltzmann equation.

In quantum mechanics, however, the notion of a phase-space
distribution directly contradicts one of its fundamental
principles: the momentum and position of a particle cannot be
determined simultaneously. Nonetheless, over the years, physicists
have introduced various quantum phase-space distributions which
reduce to $f(\vec{r},\vec{p})$ in the classical limit. These
distributions have proven extremely useful in wide-spread areas
like heavy-ion collisions, quantum molecular dynamics, signal
analysis, quantum information, non-linear dynamics, optics, image
processing, etc. One of the most frequently used is the Wigner
distribution \cite{wigner}
\begin{equation}
  W(x,p) = \int \psi^*(x-\eta/2)\psi(x+\eta/2) e^{ip\eta}d\eta \ ,
\end{equation}
where $\psi(x)$ is a quantum wave function ($\hbar=1$).
When integrating out the coordinate $x$, one gets the momentum density
$|\psi(p)|^2$, and when integrating out $p$, the coordinate space density
$|\psi(x)|^2$ follows. For arbitrary $p$ and $x$, the Wigner distribution
is not positive definite and does not have, strictly speaking, a probability
interpretation. However, there is no doubt that it contains much more
information than the usual quantum mechanical observables.
Modifed Wigner distributions with positivity are known in the literature
and are better interpreted as densities, but we leave them for a further study.

In this paper, we introduce the quark and gluon Wigner-type
phase-space distributions in hadrons such as the proton. After
certain phase-space reductions, these distributions reduce to the
generalized parton distributions (GPD's) and the
transverse-momentum dependent parton distributions which are
measurable in high-energy experiments. [For reviews of these
distributions and experiments to measure them, see \cite{review,
review1}.] Their relation to the GPD's reveals a new
interpretation of the latter---the Fourier transformation of GPD's
yields full-3D quark and gluons images of the proton at every
fixed Feynman momentum $x$, like what's seen through momentum (or
better yet ``color" or $x$-) filters. The skewness parameter $\xi$
in GPD's now has a natural interpretation: a measure of the proton
deformation along the direction selected by high-energy probes.
From the sign variation of the images, we might have a practical
way to determine how ``classical" the quarks and gluons are in the
proton.

Disregarding the renormalization scale $\mu$, the GPD's depend on
three kinematic variables $x$ (Feynman momentum), $\xi$
(skewness), and ${q}^2$ ($t$-channel momentum transfer, also
called $t$). Introduced as quantum interference amplitudes, their
original interpretation as parton and their angular momentum
densities was limited to $t=\xi=0$ \cite{ji,angmom}. In a seminal
paper by M. Burkardt \cite{burkardt}, the probability
interpretation was successfully extended to the GPD's at $t\ne
0$: the Fourier transformation of the parton densities in the
two-dimensional transverse plane (impact-parameter space). A
number of subsequent papers have studied the physical significance
of the $\xi\ne 0$ case \cite{others}, without seeking a classical
or quasi-classical interpretation. The present work expands the
2D-pictures of Burkardt along the longitudinal direction,
relating the GPD's to full-3D distributions at a fixed light-cone
momentum, interpretable as densities in the classical limit.

We begin by introducing the Wigner distribution operator for quarks in QCD
\begin{equation}
   \hat {\cal W}_\Gamma(\vec{r},k) = \int \overline{\Psi}(\vec{r}-\eta/2)
    \Gamma\Psi(\vec{r}+\eta/2) e^{ik\cdot \eta} d^4\eta \ ,
\end{equation}
where $\vec{r}$ is the quark phase-space position and $k$
the phase-space four-momentum. $\Gamma$ is a Dirac matrix
defining the types of quark denstities. $\Psi$ is a {\it gauge-invariant}
quark field in non-singular gauges (gauge potentials vanish at the spacetime infinity
\cite{jiyuan}),
\begin{equation}
        \Psi(\eta)= \exp\left(-ig\int^\infty_0 d\lambda n \cdot A(\lambda n+\eta)\right) \psi(\eta) \ ,
\end{equation}
where $n^\mu$ is a constant four-vector. Thus $\hat {\cal W}_\Gamma$
is gauge-invariant but depends on a choice of $n^\mu$.
The actual Wigner distribution can be define as the expectation
value of $\hat {\cal W}_\Gamma$ in the
hadron states. For example,
\begin{eqnarray}
     {W}_\Gamma(\vec{r}, k) &=& \frac{1}{2M}
\int \frac{d^3\vec{q}}{(2\pi)^3} \left\langle \vec{q}/2
    \left|\hat {\cal W}(\vec{r}, k)
             \right|-\vec{q}/2\right\rangle  \\
                &=& \frac{1}{2M} \int \frac{d^3\vec{q}}{(2\pi)^3}
                    e^{-i\vec{q}\cdot\vec{r}}
                   \left \langle \vec{q}/2\left|
          \hat {\cal W}(0,k)\right|-\vec{q}/2\right\rangle \ ,\nonumber
\end{eqnarray}
where $M$ is the hadron mass and the states are normalized
relativistically. The initial and final hadrons are taken with
different center-of-mass momenta, otherwise translational
invariance results in a trivial dependence on the phase-space
position $\vec{r}$ \cite{sachs}. For non-relativistic quantum
systems, the above definition readily reduces to the known Wigner
distribution. For relativistic systems such as the proton, the
above interpretation as a density neglects relativistic effects
when the momenta are comparable to the proton mass
\cite{burkardt}. Fortunately, since the proton mass is of order 1
GeV and charge radius 0.86 fm, the relativistic correction is
about \%10.



For the proton, because of the high energy involved in experiments
and because of the color confinement, the $k^-=(k^0-k^z)/\sqrt{2}$
energy distribution is difficult to measure, where the $z$-axis
refers to the momentum direction of a probe. Moreover, the
leading observables are associated with the ``good" components of
the quark (gluon) fields in the sense of light-cone quantization
\cite{lcq}, which can be selected by $\Gamma=\gamma^+$,
$\gamma^+\gamma_5,$ or $\sigma^{+\perp}$ where
$\gamma^+=(\gamma^0+\gamma^z)/\sqrt{2}$. The direction of the
gauge link, $n^\mu$, is then determined by the trajectories of
high-energy partons traveling along the light-cone $(1,0,0,-1)$.
Therefore, from now on, we restrict ourselves to the reduced
Wigner distributions integrating over $k^-$,
\begin{equation}
    W_{\Gamma}(\vec{r},\vec{k})
    = \int \frac{dk^-}{(2\pi)^2} {\cal W}_{\Gamma}(\vec{r},k) \ ,
\label{wigner}
\end{equation}
with a light-cone gauge link. Eq. (\ref{wigner}) defines the most
general (master) phase-space quark distributions relevant in
high-energy processes. Unfortunately, there is no known
experiment at the moment capable of measuring them in the full
6-dimensional phase space.

However, certain phase-space reductions of the distributions
are measurable through known processes. For example, integrating out
the transverse momentum of partons, we find
\begin{eqnarray}
        \tilde f_\Gamma(\vec{r},k^+) &=& \int \frac{d^2\vec{k}_\perp}
   {(2\pi)^2}~ W_{\Gamma}
               (\vec{r},\vec{k}) \nonumber \\
                   & =& \frac{1}{2M} \int \frac{d^3\vec{q}}{(2\pi)^3}
              e^{-i\vec{q}\cdot\vec{r}}  \int
          \frac{d\eta^-}{2\pi} e^{i\eta^-k^+}  \nonumber \\
     &&  \times \left\langle \vec{q}/2\left|\overline\Psi(-\eta^-/2) \Gamma\Psi(\eta^-/2)
\right|-\vec{q}/2\right\rangle     \ .
\end{eqnarray}
The matrix element under the integrals is
what defines the GPD's. More precisely, if one replaces $k^+$ by Feynman
variable $xP^+$ ($P^+=E_q/ \sqrt{2}$, proton energy
$E_q=\sqrt{M^2+\vec{q}^2/4}$) and $\eta^-$ by $\lambda/P^+$,
the reduced Wigner distribution becomes the Fourier transformation
of the GPD $F_\Gamma(x, \xi, t)$
\begin{equation}
     f_{\Gamma}(\vec{r},x) = \frac{1}{2M}\int \frac{d^3\vec{q}}{(2\pi)^3} e^{-i\vec{q}\cdot\vec{r}}
      F_\Gamma(x, \xi, t) \ .
\label{dis}
\end{equation}
In the present context, the relation between kinematic variables are
$\xi=q_z/(2E_q)$ and $t=-\vec{q}^{~2}$. Taking
$\Gamma = \sqrt{2}\gamma^+$, the corresponding GPD has the expansion \cite{ji}
\begin{eqnarray}
    && F_{\gamma^+}(x, \xi, t)\nonumber \\
 &=& \int \frac{d\lambda}{2\pi} e^{i\lambda x}
               \left\langle \vec{q}/2\left|\overline{\psi}(-\lambda n/2)
                  {\cal L}\sqrt{2}\gamma^+
            \psi(\lambda n/2) \right|-\vec{q}/2\right\rangle \nonumber\\
     &=& H(x, \xi, t) \overline{U}(\vec{q}/2)\sqrt{2}\gamma^+U(-\vec{q}/2) \nonumber \\
   && ~~ + E(x, \xi, t) \overline{U}(\vec{q}/2)\frac{i\sigma^{+i}q_i}{ \sqrt{2}M}
       U(-\vec{q}/2) \ ,
\label{GPD}
\end{eqnarray}
where ${\cal L}$ is the shorthand for the light-cone gauge link.

The distribution $f_{\gamma^+}(\vec{r},x)$ can be interpreted as
the 3D density in the rest frame of the proton, for the quarks
with a selected light-cone momentum $x$. Integrating over the $z$
coordinate, the GPD's are set to $\xi\sim q^z=0$, and the
resulting two-dimensional density $f_{\gamma^+}(\vec{r}_\perp,x)$
is just the impact-parameter space distribution in the Burkardt
paper \cite{burkardt}.

To understand better the physical content of the density, let us
examine its spin structure. Working out the matrix element
in Eq. (\ref{GPD}),
\begin{eqnarray}
     \frac{1}{2M} F_{\gamma^+}(x, \xi, t) = \left[H(x,\xi,t) - \tau E(x,\xi,t)\right]&
 \nonumber \\
          + ~ i (\vec{s}\times \vec{q})^z \frac{1}{2M}\left[H(x,\xi,t)+E(x,\xi,t)\right]& ,
\end{eqnarray}
where $\tau = (1+\xi)\vec{q}^2/4M^2 + \xi$. The first term is independent
of the proton spin, and the related quark density is
\begin{equation}
     \rho_+(\vec{r},x) =
   \int \frac{d^3\vec{q}}{(2\pi)^3} e^{-i\vec{q}\cdot\vec{r}} [H(x,\xi,t)-
  \tau E(x,\xi,t)]   \ .
\end{equation}
Just like charge density, it is not positive-definite at any $x$.
The second term depends on the proton spin, the corresponding
density is physically the 3rd-component of a current
\begin{eqnarray}
     j_{+}^z(\vec{r}, x) &=&  \int \frac{d^3\vec{q}}{(2\pi)^3} e^{-i\vec{q}\cdot\vec{r}}
         i (\vec{s}\times \vec{q})^z \nonumber \\
     && \times \frac{1}{2M}\left[H(x,\xi,t)+E(x,\xi,t)\right] \ .
\end{eqnarray}
The $H$-term current generates the proton's Dirac moment, and the
$E$-term generates a convection current due to the orbital
angular momentum of massless quarks and vanishes when all quarks
are in the $s$-orbit\cite{ma}. The physics in separating
$f_\gamma^+$ into $\rho_+$ and $j_+^z$ can be seen from the Dirac
matrix $\gamma^+$ selected by the high-energy probes, which is a
combination of time and space components. Because the current
distribution has no spherical symmetry, the quark density seen in
the infinite momentum frame, $\rho_++j_+^z$, is deformed in the
impact parameter space \cite{burkardt1}. This is the kinematic
effect of Lorentz transformations.

We can also study the $x$-moments of the quark and current densities.
If we integrate over $x$ directly, we obtain the electric
charge and current densities,
\begin{eqnarray}
   \rho^e_+(\vec{r}) &=& \int dx \rho_+(\vec{r}, x) \nonumber \\
&=& \int \frac{d^3\vec{q}}{(2\pi)^3}
 e^{-i\vec{q}\cdot\vec{r}} (F_1(q^2)-\tau F_2(q^2)) \ ,
   \nonumber \\
   \vec{j}(\vec{r})
&=& \int \frac{d^3\vec{q}}{(2\pi)^3}
 e^{-i\vec{q}\cdot\vec{r}} i (\vec{s}\times \vec{q}) \frac{1}{2M} (F_1(q^2) +
F_2(q^2)) \nonumber \ .
\end{eqnarray}
Notice that $\rho^e_+$ is not spherically symmetric due to the
$\xi$ dependence in $\tau$, reminiscent
of the $\gamma^+$-probe. The electric current density
has a donut shape and is responsible for the magnetic moment
of the proton \cite{sachs}. Indeed, if we calculate the proton's magnetic moment
using the classical formula, $\vec{\mu} = (1/2)\int \vec{r}\times\vec{j}$,
the right answer follows. The current generated by the $F_2(q^2)$ term
is directly related to the orbital motion of the quarks \cite{f2}.

Forming the second moment of $x$, we have the
{\it mass distribution} of the proton from the quarks \cite{note},
\begin{eqnarray}
  \rho^m_+(\vec{r}) &=& \int dx x \rho_+(\vec{r}, x) \nonumber \\
&=& \int \frac{d^3\vec{q}}{(2\pi)^3}
 e^{-i\vec{q}\cdot\vec{r}} \left[(A_q(q^2)+\xi^2C_q(q^2))\right.
\nonumber \\
&& ~~~~~~~-\left.\tau (B_q(q^2)-
\xi^2C_q(q^2)\right] \ ,
\label{gravitonm}
\end{eqnarray}
where $A_q(q^2)$, $B_q(q^2)$ and $C_q(q^2)$ are the quark
part of the proton's energy-momentum
form factors \cite{pagels,ji},
\begin{eqnarray}
  \langle P'|T^{\mu\nu}|P\rangle
  &=& \overline{U}(P')\left[A(q^2)\overline{P}^\mu \gamma^\nu
  + B(q^2) \frac{i\sigma^{\mu\alpha}q_\alpha P^\nu }{2M}\right. \nonumber \\
 && \left. + C(q^2)(q^\mu q^\nu-g^{\mu\nu}q^2)\right] U(P) \ .
\end{eqnarray}
Here the energy-momentum tensor $T^{\mu\nu}$ is symmetric and
conserved, and contain both quark and gluon contributions. The
normalization of the form factors $A(0)$ and $B(0)$ are
well-known: $A(0)=1$ is related to the energy-momentum
conservation and $B(0)=0$ is a result of the angular momentum
conservation \cite{okun,ji}.

The energy-momentum form factors can {\it in principle} be
measured in elastic graviton-nucleon scattering. Here we use this
fact to motivate the physical interpretation of
$\rho^m_+(\vec{r})$ as the proton's mass distribution. Gravitions
have spin 2, and an on-shell massless graviton has helicity
$\lambda=\pm 2$. An off-shell (massive) graviton can have
additional helicity states $\lambda=\pm 1$ and $0$. Consider the
graviton-proton scattering in the Breit frame. Assume the helicity
of the initial proton is 1/2, then the final proton has helicity
$1/2-\lambda$. Since the proton has spin-1/2, only off-shell
gravitons of helicity $1$ and $0$ contribute. A graviton with
helicity $1$ and $0$ acts very much like a photon, and we can
mimic the Sachs's electromagnetic form factors by defining the
gravitational ``electric" (or mass) form factor
\begin{equation}
       G_{E}^m(q^2) = A(q^2) + \frac{t}{ 4M^2}B(q^2) \ ,
\end{equation}
proportional to the helicity non-flip graviton-nucleon scattering
amplitude \cite{jisong,belitsky}. If one ignores the deformation
due to the probes ($\xi=0$), $\rho^m_+(\vec{r})$ in Eq.
(\ref{gravitonm}) is just the Fourier transformation of the above
form factor.

The $x$-moment of the quark current is the {\it momentum density}
in the proton
\begin{eqnarray}
    j^{zp}(\vec{r})
   &=& \frac{M}{2}\int dx x j_{+}^z(\vec{r},x) \nonumber \\
   &=& \int \frac{d^3\vec{q}}{(2\pi)^3}
 e^{-i\vec{q}\cdot\vec{r}} i (\vec{s}\times \vec{q})^z \nonumber \\
 && \times \frac{1}{4M} \left[A_q(q^2)+ B_q(q^2)
\right] \ .
\end{eqnarray}
The combination in the integrand is the gravitational ``magnetic"
(or angular momentum) form factor,
\begin{equation}
      G_{M}^m(q^2) = A(q^2) + B(q^2) \ ,
\end{equation}
proportional to the helicity-flip graviton-proton scattering amplitude
\cite{jisong,belitsky}.
Knowing the momentum density as a function of $\vec{r}$, it is simple to calculate
the quark contribution to the angular momentum (spin) of the proton according to
$\int \vec{r}\times \vec{j}^p$. The answer is
$(A_q(0)+B_q(0))/2$ \cite{ji}.  A contribution from the gluons
can be obtained in a similar way.


Let us remark that gravity plays a dominant role at two extreme
distance scales: the cosmic scale and the Plank scale. At the
atomic and subatomic levels, with the exception of the motion of
these particles under strong gravitational fields, gravitation has
little effects because of its extremely weak coupling. Thus while
mass distribution and moment-of-inertia are important concepts in
classical mechanics, they have rarely been discussed for
microscopic quantum systems such as the proton. Thus, it is very
interesting that the GPD's provide gravitational form factors
without actual gravitational scattering \cite{ji,jisong}. A study
of the gravitational Sachs's form factors (and gravitational
radius) in chiral perturbation theory can be found in
\cite{belitsky}. A related coordinate-space interpretation of the
gravitational form factors, in particular, the stress density
$T^{ij}$, can be found in Ref. \cite{polyakov}.

If integrating over $\vec{r}$ in the Wigner distributions in
Eq. (\ref{wigner}), one obtains the transverse-momentum
dependent parton distributions. There is a lot of interesting physics
associated with these distributions which has been discussed recently.
For instance, in a transversely polarized proton, the quark
momentum distribution has an azimuthal angular dependence
\cite{sivers,review1,jiyuan}. The so-called
Siver's function can produce a novel single-spin asymmetry in
deep-inelastic scattering. We will not pursue this topic here,
except emphasizing that they have the same generating
functions as the GPD's.

The phase-space distributions in the transverse coordinates
have also been used in small-$x$ physics \cite{muller}, where
the parton physics becomes truly classical at
high density.

We end the paper by making a few remarks. First, as we have
mentioned before, the Wigner distributions are quantum
phase-space distributions which contain much more information of
a quantum system than conventional observables. Originally
introduced as a theoretical construction, it is surprising that
these distributions for the proton are directly measurable
through high-energy processes. Second, the Wigner distributions
are more than just quantum interference amplitudes; they have a
clear physical interpretation in the classical limit. This endows
the GPD's a new physics feature which has not been recognized
before. QCD in the limit of large number of colors might be a
case in which the Wigner distributions become truly classical
densities. Finally, the extent to which the Wigner distributions
in the proton can be regarded as classical densities may be judged
once the phenomenological GPD's is available from experiment data.

I first heard a possible connection between GPD's and Wigner
distribution from A. Belitsky, and this work is inspired by him
and T. Cohen. I also thank M. Burkardt for many helpful
discussions about his impact-parameter space interpretation of
the GPD's, and M. Diehl for critical comments. This work was
supported by the U. S. Department of Energy via grant
DE-FG02-93ER-40762.

\end{document}